# Hidden Markov Model Based Approach for Diagnosing Cause of Alarm Signals

**Joshiba Ariamuthu Venkidasalapathy[a,b]. Costas Kravaris[a,*]**

[a]Artie McFerrin Department of Chemical Engineering, Texas A&M University, College Station, TX 77840, USA

[b]Mary Kay O'Connor Process Safety Center, Texas A&M University, College Station, TX 77840, USA

## *Abstract*

When a fault occurs in a process, it slowly propagates within the system and affects the measurements triggering a sequence of alarms in the control room. The operators are required to diagnose the cause of alarms and take necessary corrective measures. The idea of representing the alarm sequence as the fault propagation path and using the propagation path to diagnose the fault is explored. A diagnoser based on hidden Markov model is built to identify the cause of the alarm signals. The proposed approach is applied to an industrial case study: Tennessee Eastman process. The results show that the proposed approach is successful in determining the probable cause of alarms generated with high accuracy. The model was able to identify the cause accurately, even when tested with short alarm sub-sequences. This allows for early identification of faults, providing more time to the operator to restore the system to normal operation.

______________________________

*corresponding author

E-mail address: Kravaris@tamu.edu

Contact number: +1 979-458-4514

1. INTRODUCTION

Process variables are continuously measured and monitored in processing plants to ensure safe and efficient operation. There are preset normal operating limits that are defined for the measured variables which will be the basis of classifying the operation as normal or abnormal. An abnormal scenario or fault refers to a condition when the measurements deviate from these normal operating limits. Under such a scenario, an alarm is triggered in the control room in the form of a light or sound to alert the operator. The common practice is to generate alarms based on deviation in a single measurement, also known as univariate process monitoring. The operator is provided with guidelines consisting of corrective actions to be performed in case a measurement alarm is triggered. An alarm system comprises of both the measurement signals and guidelines for operators. It is a signal-based fault detection and diagnosis system that notifies the operator of a process anomaly and requires an operator to perform corrective action. Alarm systems play an important role in process safety since it is one of the first layers of protection to detect an abnormal scenario and prevent an incident form happening. In [1], it is shown that there is strong evidence that human operators still play a crucial role in control of processes, in spite of improved automation and advanced process control systems. Hence, it is important that operators are not merely overloaded with information from alarm system but provided with possible cause of alarms that will lead to taking the correct response measure to restore the system to normal operation.

Incident investigation reports reveal that inefficient alarm system design has been an important contributing factor in incidents such as Three Mile Island Nuclear plant disaster, BP Texas city refinery explosion incident and Texaco Pembroke Refinery accident[2, 3]. General guidelines for alarm management were published by the Engineering Equipment and Materials User's Association[4] and ISA 18.2 alarm management lifecycle[5]. These guidelines provide recommendations and requirements on the alarm system key performance indicators. However, there are no systematic procedures for attaining acceptable alarm system performances. In addition to these, several researchers have studied different aspects of alarm management that includes design of alarm limits for variables, prioritization of alarms[6], reducing the number of alarms[7-9], variable selection[8, 10] and identifying the cause of alarms[11-15].

The common practice in industries is to map each alarm signal to a list of possible causes and corrective actions that are established with the help of qualitative cause-effect based techniques such as HAZOPs and HAZIDs. These guidelines may be sufficient for the operators to troubleshoot small disturbances and process deviations. However, they are not sufficient to capture the complex interactions between process variables particularly during a major abnormal event that may generate a number of alarms. Recently, several advancements have been made to address these challenges in designing and managing efficient alarm systems. Particularly, the big data era has paved way for exploring data-driven methods for improving the performance of alarm systems[16].

The data driven methods with applications in chemical systems typically use statistical analysis methods such as principal component analysis (PCA)[17-22], partial least squares (PLS)[21, 23], independent component analysis (ICA)[24-26] hidden Markov model[27] and Fisher discriminant analysis (FDA). Machine learning approaches for FDD with chemical process systems applications are primarily based on artificial neural networks[28], neuro fuzzy methods[29, 30], support vector machine (SVM)[31, 32], k-nearest neighbor (knn)[33, 34] and Bayesian network (BN)[22, 35-38]. Most of these methods utilize continuous process variable measurements as the input to the model. In this work, the objective is to utilize the information from the alarm signals as the input to reduce the computational complexity of the FDD tool and to make it fairly interpretable to the operator.

There have also been other data-driven methods present in literature addressing the problem of diagnosing the cause of alarm signals using the binary alarm systems information. These alarm sequence classification techniques can be broadly categorized as sequence alignment based classification, feature based classification, root cause analysis based classification and discrete event system diagnosis using Petri nets. The sequence alignment based approaches aim to capture the similarity between sequences to determine the fault[11, 12, 39, 40]. Feature based approaches transform a sequence into a feature vector and then apply conventional classification methods such as k-nearest neighbours (knn), neural networks or support vector machines[13, 41]. One of the limitations of the sequence alignment and feature based approaches is their requirement to transform a sequence into a feature vector which can be nontrivial[42]. In root cause analysis methods, transfer entropy and Bayesian networks have been used to capture indirect relationship between process variables from alarm system database[14, 43, 44]. Other methods include the use of Petri nets [45] to build a model and algorithm for fault management. Although these methods identify the indirect causal relationship between variables, they have not been used as a diagnosing tool to help the operator troubleshoot alarm cause in real time.

Another approach commonly used in classification of sequences are the statistical model based methods. A simple model based diagnoser to identify the cause of alarm signals using only the binary alarm sequence information has not been found in literature. In model based approaches, probability measures of output sequences in each class are represented using statistical models such as Naive Bayes sequence classifier, Markov model or hidden Markov model (HMM). HMM has been used in FDD techniques in the past, both in the context of fault detection and for fault isolation[25, 46-48]. However, an HMM based approach that uses only the binary alarm signal information for isolating the fault has not been studied in the past. The motivation is to develop a simple diagnoser that is easy to train, computationally inexpensive and quickly isolates the fault providing operators enough time to respond to the situation.

In this paper, an HMM based approach is proposed for alarm sequence classification. The probabilistic framework of HMM helps capture the stochastic elements present in realistic plant operations such as measurement noises, fault characteristics and external disturbances. The model parameters of HMM (which will be discussed in detail in the next section) are fairly interpretable in nature and provide scope to integrate failure rates of equipment to improve classification accuracy. The proposed approach integrates information from safety review processes, process dynamic simulator and alarm system database to build a robust tool for diagnosing the cause of alarm signals as shown in Figure 1.

The paper is organized as follows: in section 2 a brief overview of HMM and the approach proposed for diagnosing cause of alarm signals is outlined. The approach is applied to the Tennessee Eastman process and the results are compared to another binary alarm signals based classifier in section 3. Lastly, the conclusions are presented in section 4.

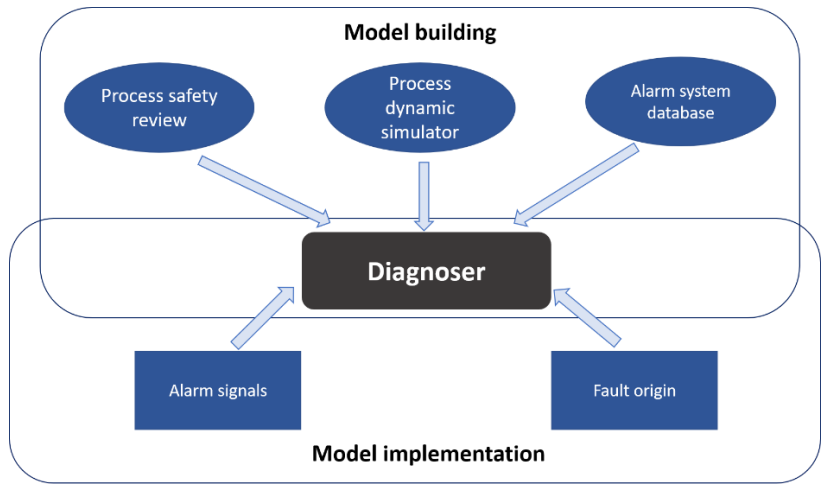

Figure 1: Schematic of the diagnoser

2. APPROACH PROPOSED FOR DIAGNOSING ALARMS

*2.1 Brief overview of HMM*

HMM is a probabilistic model that assumes the states follow the Markov property and are unobservable[49]. The outputs that depend on the state are observable. The theory was first published in[50, 51]. Their applications include speech recognition, facial expression identification and bioinformatics. Refer to Table 1 below for the list of notations used in this section along with its description.

Table 1. List of notations used in section 2 and their description

| Notation | Description |
|---|---|
| $N$ | Number of distinct states |
| $S_i$ | State i, i= 1, …, N |
| $t$ | Time instants |
| $q_t$ | Actual state at time t |
| $a_{ij}$ | Probability of transitioning from state i to state j<br>i= 1,…, N, j= 1,…, N |
| $A = \{a_{ij}\}$ | State transition probability matrix |
| $M$ | Number of measurements |
| $v_k$ | Measurement i, k = 1,…, M |
| $O_t$ | Actual measurement at time t |
| $b_j(k)$ | Probability of emitting measurement $v_k$ from state $S_j$ |
| $\pi_i$ | Initial state distribution, the probability that state $S_i$ is at time 0 |
| $\lambda$ | Denotes an HMM, $\lambda = (A, B, \pi)$ |
| $O_{0:T} = O_0, O_1, O_2, O_3, \ldots O_T$ | An observation sequence |
| $\alpha_k(q_k)$ | The joint probability distribution of states at time k given the observables sequence up to time $k$ |
| $\beta_k(q_k)$ | The conditional probability of the observed data from time $k+1$ given the state at time $k$ |
| $\eta_k(q_k) = P(q_k\vert O_{0:T})$ | The joint probability distribution of a state at time $k$ given $O_{0:T}$ |
| $\xi_k(q_k, q_{k+1}) = P(q_k, q_{k+1}\vert O_{0:T})$ | The joint probability of observing two consecutive states given $O_{0:T}$ |
| $Q = q_{0:T} = q_0, q_1, q_2, q_3, \ldots q_T$ | The hidden state sequence |
| $\delta_k(i)$ | the highest probability along a single path at time $k$ that accounts for the first $k$ observables and ends in state $S_i$ |
| $\psi_k(i)$ | A variable in the Viterbi algorithm |

In discrete HMM, a system is described as being in one of a set of $N$ distinct states, $S_1, S_2, \ldots, S_N$ at any time. More details on other types of HMMs can be seen in[52]. The states of the system evolve form one state to another according to the transition probabilities associated with the state. The time instants associated with the state changes can be denoted by $t = 1,2,\ldots$, and the actual state at time $t$ as $q_t$. For first order Markov chain, the probabilistic description is truncated to just the current and the predecessor state. It can be represented as below.

$$P(q_t = S_j | q_{t-1} = S_i, q_{t-2} = S_k, \ldots) = P(q_t = S_j | q_{t-1} = S_i) \quad (1)$$

The states of the systems are often not observable in realistic scenarios; hence to make a Markovian process less restrictive, the HMM allows modelling the observables as a probabilistic function of the state. The resulting model is a doubly stochastic process with an underlying stochastic process that can be observed through another set of stochastic processes that produce the sequence of observations.

The model parameters include the following: the number of distinct states $N$, the state transition probability matrix $A$, the number of measurement variables at each state $M$, the emission probability matrix $B$ and the probability distribution of the initial state $\pi$. The $N$ states can be denoted by $S = \{S_1, S_2, \ldots, S_N\}$, the $M$ distinct observables from the states by $V = \{v_1, v_2, \ldots, v_M\}$. The hidden state and the measured output at time t are represented by $q_t$ and $O_t$ respectively. The state transition probability matrix $A_{ij} = \{a_{ij}\}$, where $a_{ij}$ is the probability that the state evolves from state $S_i$ to state $S_j$.

$$a_{ij} = p(q_t = S_j | q_{t-1} = S_i), (1 \leq i, j \leq N) \quad (2)$$

The transition probabilities also satisfy $a_{ij} \geq 0$ and,

$$\sum_{j=1}^{N} a_{ij} = 1 \qquad (3)$$

The emission probability distribution in each state $B = \{b_j(k)\}$ is the probability of observing output $v_k$ from state $S_j$.

$$b_j(k) = p(O_t = v_k | q_t = S_j), (1 \leq j \leq N, 1 \leq k \leq M) \qquad (4)$$

The initial state distribution $\pi_i$ is the probability that the model is at state $S_i$ at $t = 0$. It is given by,

$$\pi_i = p(q_0 = S_i), (1 \leq i \leq N) \qquad (5)$$

To summarize, a complete specification of an HMM requires defining two model parameters $N$ and $M$ and the specification of the observation symbols, and the specification of the three probability measures $A, B$ and $\pi$. An HMM can be denoted by,

$$\lambda = (A, B, \pi). \qquad (6)$$

There are three basic problems for HMMs [52]. The first is to compute the probability $P(O_{0:T}|\lambda)$ of a given observation sequence $O_{0:T} = O_1, O_2, O_3, \ldots O_T$ and the model $\lambda = (A, B, \pi)$. The second problem aims to identify the hidden state sequence corresponding to the given observation sequence $O_{0:T}$ and model $\lambda$, which best explains the observations for some optimality criteria. The third problem is to optimize the model parameters so as to maximize $P(O_{0:T}|\lambda)$ for the given observation sequence.

*2.2 Summary of algorithms for HMM learning and decoding of hidden state sequence*

In this section, algorithms available for model learning and decoding hidden state sequence are briefly summarized. The Baum-Welch algorithm[50, 51] is commonly used for learning the HMM parameters. The Viterbi algorithm[53, 54] is commonly used for identifying the most likely hidden state sequence given an observation sequence.

*Baum-Welch algorithm for training the HMM model λ*

The Baum-Welch algorithm is also known as the forward-backward algorithm. In the initial step, the parameters $A, B$ and $\pi$ are initialized. They are randomly assigned if there is no prior knowledge. Given the initial model parameters and the observed sequence $O_{0:T}$, the following steps are performed. In the forward step, $\alpha_k(q_k)$, which is the joint probability distribution of states at time k given the observables sequence up to time $k$, is calculated.

$$\alpha_k(q_k) = P(O_{0:k}, q_k) = \sum_{q_{k-1}=1}^{n} \alpha_{k-1}(q_{k-1})P(q_k| q_{k-1})P(O_k| q_k) \qquad (7)$$

$$\alpha_0(q_0) = P(O_0, q_0) = P(O_0| q_0)P(q_0) \qquad (8)$$

In the backward step, $\beta_k(q_k)$, which is the conditional probability of the observed data from time $k + 1$ given the state at time $k$, is calculated.

$$\beta_k(q_k) = P(O_{k+1:T}, q_k) = \sum_{q_{k+1}=1}^{n} \beta_{k+1}(q_{k+1})P(q_{k+1}| q_k)P(O_{k+1}| q_{k+1}) \qquad (9)$$

$$\beta_T(q_T) = 1 \qquad (10)$$

The forward and backward steps are then combined to calculate the joint probability distribution of a state at time $k$ given $O_{0:T}$.

$$\eta_k(q_k) = P(q_k| O_{0:T}) = \frac{\alpha_k(q_k)\,\beta_k(q_k)}{\sum_{q_k=1}^{n} \alpha_k(q_k)\beta_k(q_k)} \qquad (11)$$

The joint probability of observing two consecutive states given $O_{0:T}$ is also calculated.

$$\xi_k(q_k, q_{k+1}) = P(q_k, q_{k+1}|O_{0:T}) \quad (12)$$

$$\xi_k(q_k, q_{k+1}) = \frac{\alpha_k(q_k)\beta_{k+1}(q_{k+1})P(q_{k+1}|q_k)P(O_{k+1}|q_{k+1})}{\sum_{q_k=1}^{n} \alpha_k(q_k)\beta_{k+1}(q_{k+1})P(q_{k+1}|q_k)P(O_{k+1}|q_{k+1})} \quad (13)$$

Finally, in the update step, the model parameters are updated to maximize the probability of obtaining the observed sequence.

$$\pi_0^* = \eta_0(q_0) \quad (14)$$

$$a_{ij}^* = P(q_{k=j}|q_{k-1=i}) = \frac{\sum_{k=1}^{T} \xi_k(q_{k=j}, q_{k-1=i})}{\sum_{k=1}^{T} \eta_k(q_{k-1=i})} \quad (15)$$

$$b_{ij}^* = P(O_{k=j}|q_{k=i}) = \frac{\sum_{k=1}^{T} \eta(q_{k=i}) \times 1_{O_k=j}}{\sum_{k=1}^{T} \eta_k(q_{k=i})} \quad (16)$$

*Viterbi algorithm for calculating the hidden state sequence*[53,54]

Viterbi algorithm uses a recursive algorithm to identify the single best hidden state sequence based on a trained model and the given observation sequence. The most probable hidden state sequence is defined by,

$$q_{0:T}^* = \underset{q_{0:T}}{\mathrm{argmax}}\, P(q_{0:T}|O_{0:T}; \lambda^*) \quad (17)$$

For determining the best hidden state sequence $Q = \{q_0, q_1, \ldots, q_T\}$ for the given observation sequence $O = \{O_0, O_1, \ldots, O_T\}$, the following quantity needs to be defined.

$$\delta_k(i) = \underset{q_0, q_1, \ldots, q_{k-1}}{\max}\, P(q_0, q_1, \ldots, q_k = i, O_0, O_1, \ldots, O_k|\lambda) \quad (18)$$

$\delta_k(i)$ represents the highest probability along a single path at time $k$ that accounts for the first $k+1$ observables and ends in state $S_i$.

By induction,

$$\delta_{k+1}(j) = \left[\max_i \delta_k(i) a_{ij}\right] \cdot b_j(O_{k+1}) \quad (19)$$

The argument that maximizes the above step is stored via the array $\psi_k(i)$. The algorithm consists of an initialization step, recursion step, termination, and path backtracking step.

In the initialization step,

$$\delta_0(i) = \pi_i b_i(O_0), \ 1 \leq i \leq N \quad (20)$$

$$\psi_0(i) = 0 \quad (21)$$

During recursion step,

$$\delta_k(j) = \max_{1 \leq i \leq N}[\delta_{k-1}(i) a_{ij}] b_j(O_k), \ 1 \leq k \leq T, 1 \leq j \leq N \quad (22)$$

$$\psi_k(j) = \underset{1 \leq i \leq N}{\operatorname{argmax}} [\delta_{k-1}(i) a_{ij}], 1 \leq k \leq T, 1 \leq j \leq N \quad (23)$$

At the termination step,

$$P^* = \max_{1 \leq i \leq N}[\delta_T(i)] \quad (24)$$

$$q_T^* = \underset{1 \leq i \leq N}{\operatorname{argmax}} [\delta_T(i)] \quad (25)$$

Lastly, the state sequence is determined in the backtracking step by,

$$q_k^* = \psi_{k+1}(q_{k+1}^*), k = T-1, T-2, \ldots 0 \quad (26)$$

More details on the algorithm can be seen in[52-54]

*2.3 Approach proposed for fault diagnosis based on alarm signals*

The proposed diagnoser is based on the idea that the alarm sequences generated by a fault are representative of the fault's propagation path. In other words, the alarm sequence captures how a fault propagates within the system. The propagation paths of a fault are assumed to be the signatures of that fault and they will be used to diagnose the cause of alarm signals. Throughout this paper, it will be assumed that faults occur one at a time. The sequence in which the alarms are triggered can be affected by the noises in measurement signals, the alarm settings on the measurements and the magnitude of the fault. In this paper, the diagnoser is built based on the HMM discussed in section 2.1 for alarm sequence classification. In particular, the doubly stochastic framework of the HMM allows to model the faults as the states that are unobservable and the alarm sequences as the observable outputs. While training the model, the Baum-Welch algorithm (discussed in section 2.2) captures the sequence information by tuning the model parameters A, B and $\pi$ in such a way that the probability of observing the alarm sequences in the training data are maximized. The assumption that the state (or the hidden fault) transition dynamics is governed by the Markovian process enforces the constraint that the future fault depends only on the current fault of the system. This, along with the assumption that only one fault occurs at a time, enforces the transition probability to have a structure similar to the diagonal matrix. The diagnoser uses the Viterbi algorithm to identify the single best fault sequence generated by the given alarm sequence (either from the test data or real-time data). Ideally, for a given alarm sequence, the result of the diagnoser is expected to be a single recurring state that corresponds to the fault in the system. However, due the stochastic nature of the process, the diagnoser might predict state transitions involving other faults as well. Hence, the diagnoser designates the most recurring state in the predicted state sequence as the fault.

As the first step to building the diagnoser, the potential faults or hazardous scenarios need to be identified. For this purpose, a safety review technique such as what if analysis, preliminary hazard analysis, hazard and operability study needs to be performed. More details on specific hazard identification techniques can be seen in[55, 56]. It is assumed that a list of faults generated based on the safety review processes mentioned above is available. The next step to building the diagnoser, is to generate data by simulating the identified faults with the help of a dynamic simulator. Hence, it is assumed here that a dynamic plant simulator is available for obtaining data. In fact, many processing industries use simulators for training operators to respond to emergency situations[57, 58]. Another way to obtain the alarm sequences for the identified faults is to extract them from the alarm system database.

In order to build a robust diagnoser, it is important to capture different possible fault propagation paths that arise due to differences in fault magnitudes and sensor noises. Hence, it is important to simulate a range of fault magnitudes to extract alarm sequences. Accounting for random noises that affect commercially available sensors will improve the overall performance of the diagnoser. Once alarm sequence data are obtained for all faults, either based on plant simulator or alarms database, the faults are modelled as the hidden states and alarm sequences as the observable outputs of the HMM.

The following criteria was used to extract the true fault propagation path from the training data. A measurement is considered to be affected by a fault if and only if it is in its alarm state continuously for at least a predefined time period T. The user defined parameter T is chosen based on the false alarm rate and alarm limit settings in the plant. Even if the basic process control system might have restored the measurement to normal operating limits after time T, it is considered that the measurement is affected by the fault.

Available data are split into training and test datasets to learn and validate the model, respectively. The number of fault scenarios identified in the safety review process will become the HMM parameter $N$ and the number of measurements in the process will become the HMM parameter $M$. The model parameters $A, B$ $and$ $\pi$ can be obtained using the Baum-Welch algorithm (eq.(7) – (16)) discussed in section 2.2. Industry failure rate database can be used as initial guess for the initial probability parameter $\pi$, if available. If not, all faults can be assumed to be equally likely. For validating the trained model, the Viterbi algorithm discussed in section 2.2 is used. The Viterbi algorithm calculates the most probable state sequence corresponding to the given observation and model using eq. (26). The diagnoser then identifies the most commonly occurring state in the most probable state sequence as the hidden fault. Using the Viterbi algorithm, it is also possible to determine the second most likely state sequence. The most commonly occurring state in this sequence can be called second most likely fault. Once the model is validated using the test data set, it will be used for online diagnosing of alarm signals. The diagnoser could notify the operators with the most and second most probable causes of the alarm signals if needed. The schematic of the proposed approach is shown Figure 2. The approach will be tested on a case study in the next section.

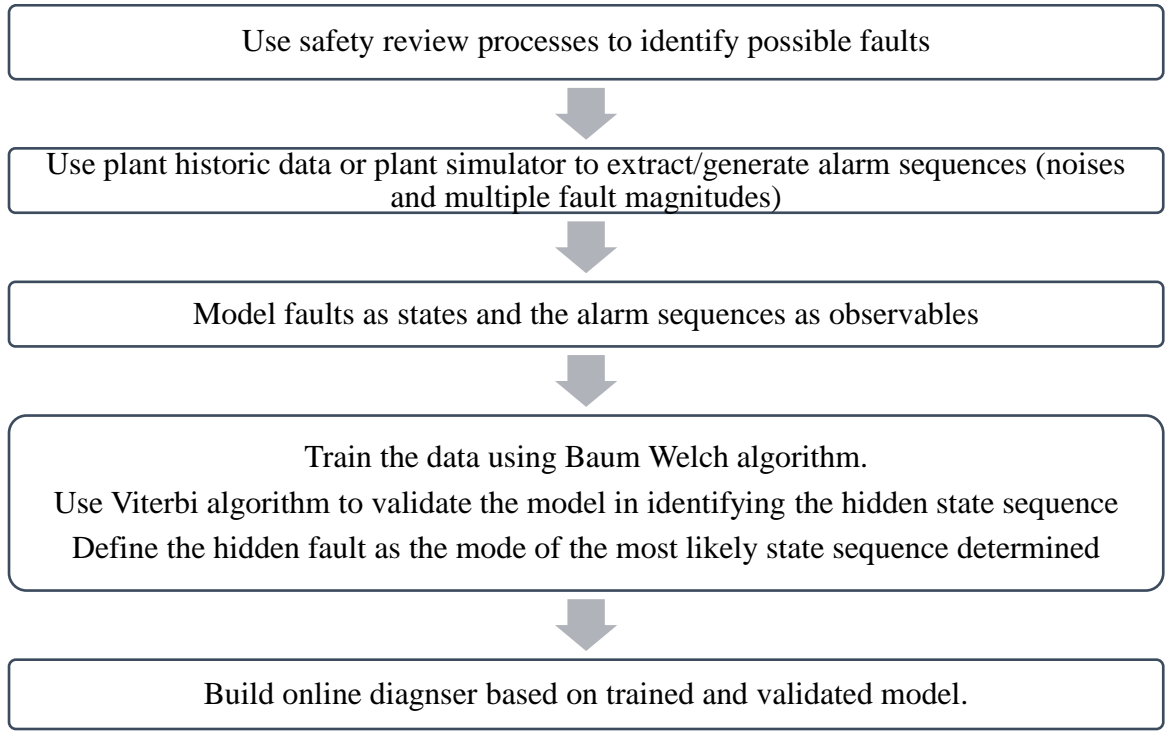

Figure 2: Schematic of the proposed approach for diagnosing cause of alarm signals

3. CASE STUDY

The proposed approach for diagnosing the alarm signals and identifying the cause of alarms using hidden Markov model is tested on an industrial case study: the Tennessee Eastman (TE) process control problem, published in [59].

3.1 Process description and fault characterization

The Tennessee Eastman process has been used by the academic community from different areas of research, including process control design, process optimization, fault diagnosis and process safety to test and validate their schemes and designs[8, 60-62]. The process consists of 41 measured variables, 50 states and 12 manipulated inputs. The Simulink model "MultiLoop-mode1. mdl" in Tennessee Eastman Challenge Archive[63], which is based on the decentralized control developed by Ricker employing PI controllers, is used to carry out the simulation of faults in this research. The closed-loop simulator terminates when any of the safety constraints identified in[59] are violated. For this case study, an alarm is set to trigger when the measured variables go beyond 3 standard deviations from the mean value measured at normal operating conditions. The parameter $T$ discussed in section 2.3 is chosen as 300 seconds for this case study.

A list of ten faults identified for the TE process are shown in Table 2. Faults list along with the number of sequences used in training and testing the diagnoser, and minimum and maximum sequence length. The faults are chosen to represent abnormal situations that results in several alarms and could result in the violation of one or more safety constraints[59] within first few hours of fault introduction. Since the disturbances in Table 8 of [59] can be handled well by the control system designed in [61], which is used for the simulation of faults, only the severe faults that result in a number of alarms are considered in this paper (refer to Table 2). The faults are modelled as step inputs of different fault magnitudes. They are identified based on valve stiction and sensor malfunctions (faults 1-10). The high and low alarm limits for measurements are distinguished as two different measurements to enhance the classification accuracy. For modelling purposes, the high alarm limits of the measurements are numbered 1- 41 and the corresponding low alarm limits are numbered 42-82. The nonlinear dynamic model was used to simulate these scenarios for different fault magnitudes. The observed alarm sequences for a fault varied in the order in which the alarms were generated as well as the length of sequences. A total of 65 fault scenarios were simulated for training the HMM parameters by varying the step size of the fault. In addition to this, 42 different fault scenarios were simulated for evaluating the performance of the diagnoser.

Table 2. Faults list along with the number of sequences used in training and testing the diagnoser, and minimum and maximum sequence length

|  |  | Training | | | Testing | | |
|---|---|---|---|---|---|---|---|
| Fault no. | Fault description | No. of seqs. | Min. seq length | Max seq. length | No. of seqs. | Min. seq length | Max seq. length |
| 1 | Reactor temperature gauge drift | 5 | 15 | 18 | 4 | 15 | 15 |
| 2 | C feed valve stuck | 8 | 12 | 37 | 4 | 11 | 31 |
| 3 | E feed valve stuck | 7 | 31 | 36 | 4 | 31 | 35 |
| 4 | D feed valve stuck | 7 | 35 | 38 | 4 | 36 | 38 |
| 5 | Reactor pressure gauge negative drift | 6 | 12 | 12 | 4 | 12 | 12 |
| 6 | Separator level gauge negative drift | 6 | 12 | 35 | 5 | 12 | 35 |
| 7 | Condenser cooling water gauge | 6 | 18 | 20 | 5 | 19 | 20 |
| 8 | A feed valve stuck | 6 | 30 | 35 | 4 | 30 | 31 |
| 9 | Purge valve stuck | 6 | 10 | 15 | 4 | 9 | 15 |
| 10 | Reactor coolant valve stuck | 8 | 15 | 35 | 4 | 28 | 36 |
|  | Total number of sequences | 65 |  |  | 42 |  |  |

3.2 Results and discussions

The faults are modelled as states and the alarm sequences are modelled as the observed output sequences as discussed before. HMM parameters were trained using the MATLAB Statistics and Machine Learning Toolbox that uses the Baum-Welch algorithm discussed in section 2.2. All states are assumed to be equally likely for being the initial state since there was no prior failure probability rate information available for this case study. The time for training the HMM was less than second (performed on an Intel Core i7- 8565U processor with 16GB RAM) that makes this approach computationally less expensive.

The Viterbi algorithm (discussed in section 2.2) in the MATLAB toolbox was used to determine the single best hidden state sequence that maximizes the probability of seeing the given observation sequence. The time for identifying the hidden sequence was less than a second (performed on an Intel Core i7- 8565U processor with 16GB RAM), which makes it highly suitable for fault diagnosis in industrial applications. The accuracy of the fault classification is shown as a function of the length of the sequence provided to the diagnoser as input in Figure 3. While plotting Figure 3, it is assumed that for a test sequence of length $l$, the diagnoser's fault estimation for any input subsequence of lengths greater than $l$ will be defined as that of the result obtained using the whole sequence (length $l$) as input. As the length of input sequence increases, the accuracy of classification increases on an average. The diagnoser could successfully identify the fault for sequence lengths greater than 12 with more than 96% accuracy. There is at least 90% accuracy or more for sequence lengths greater than 3, which shows that by merely utilizing the binary alarm sequence information, it is possible to quickly diagnose the cause of the abnormal event with the proposed diagnoser. The true fault and fault determined by the diagnoser for the 42 test sequences with the various input sequence lengths are shown in Figure 4.

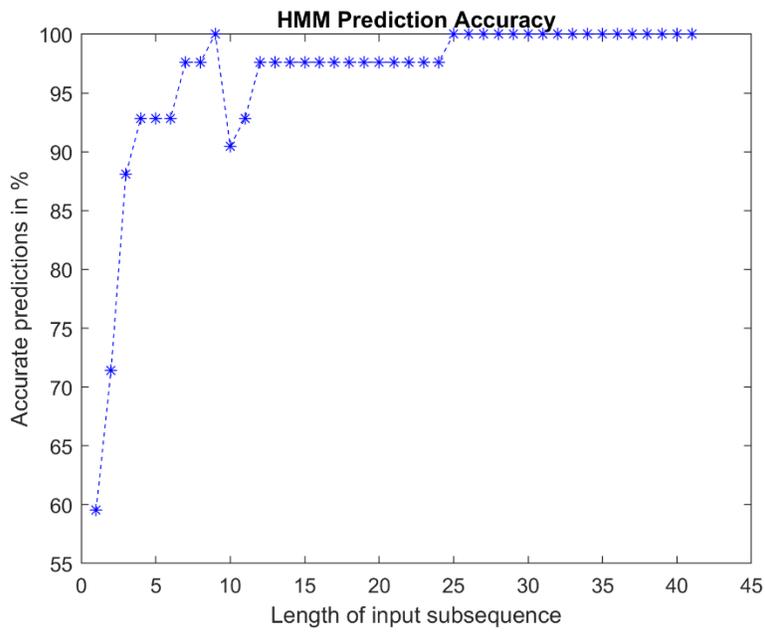

Figure 3: Length of observation sequence provided as input to the diagnoser vs accuracy in classification for the test data with 42 sequences

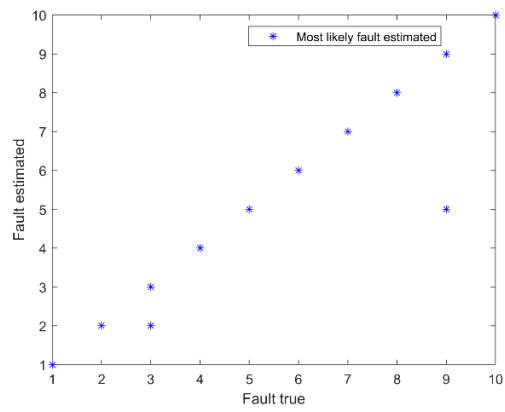

(a).

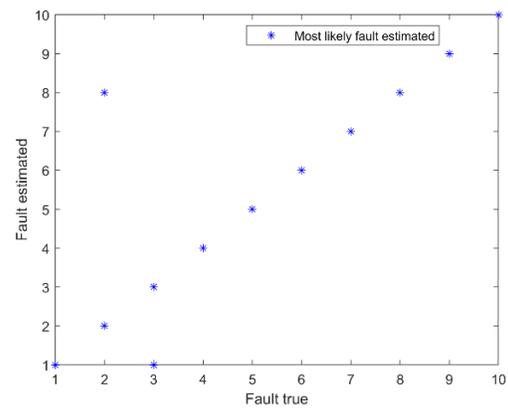

(b).

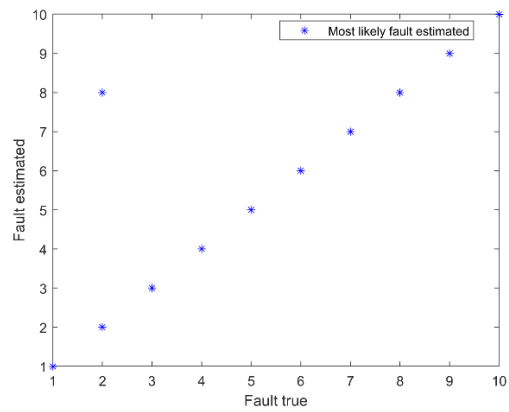

(c).

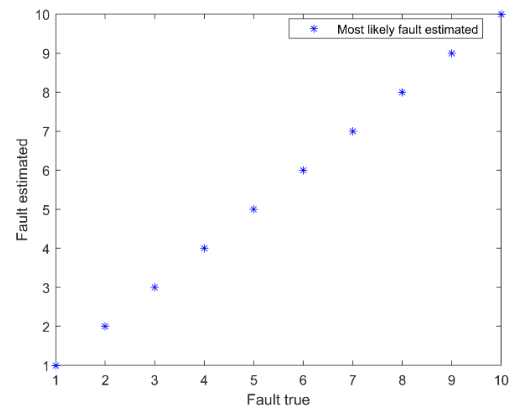

(d).

Figure 4: Fault diagnosed vs true fault for an input sequence length of (a). 6, (b). 10, (c). 12 and (d). 38

It was observed from Figure 4 that a majority of the misclassifications arose due to the diagnoser's inability to distinguish the faults from the same control loop. Faults 2 and 8 are from the same control loop and faults 5 and 9 are from the same control loop. In Figure 4. (b) and Figure 4.(c), the fault 2 is misclassified as the fault 8. In Figure 4.(a), the fault 9 is misclassified as the fault 5. This suggests a problem of distinguishability between faults within the same control loop since their fault propagation paths could be closely related. Another misclassification arose when fault 3 was classified as either fault 1 or fault 2 as seen in Figures 4. (a) and 4. (b). There was no problem distinguishing fault 3 for input sequence lengths greater than 12. The fault propagation paths for the faults 1 (reactor temperature sensor fault), 2 (feed C valve stiction) and 3 (feed E valve stiction) are shown in Figure 5 to help understand the reason for misclassification. A closer look at the Figures 5. (a) and 5. (c), show that the path captured by the first few alarms triggered seem to affect the reactor temperature at first, then move on to affect the pressure of all three vessels, then the liquid levels in the vessels, then purge rate, purge gas composition and subsequently, the temperature of the stripper and product separator vessels. In addition to these measurement alarms, the separator cooling water outlet temperature, recycle gas flowrate and feed composition are affected in the case of fault 3. Due to the high similarity observed in the first few elements of the alarm sequences of these two faults, there is a problem of distinguishability when the input sequence length is less than 12. Similarly, fault 3 has a path that is very similar to fault 2 (refer to Figures 5. (a) and 5. (b)), when only the first few elements of the alarm sequences are considered.

The proposed method was compared to the similarity approach developed in [11] that uses binary alarm sequence data as input to identify clusters of similar alarm sequences, based on frequency of consecutive alarms. Their method proposes identifying predetermined set of corrective measures for each cluster to guide operators when alarms similar to these clusters are detected. Their approach was chosen for comparison with the proposed approach for the following reasons: (i) the calculation of similarity between sequences are interpretable to a certain extent and (ii). the computational complexity was comparable to the proposed approach. Their method broadly involves: (i). eliminating the effect of chattering alarms, (ii). mapping each sequence into a feature matrix $P$, where $P_{ij}$ for $i,j = \{1,..,M\}$ ($M$ is the number of measurements) is the frequency of alarm $j$ coming immediately after alarm $i$, (iii). calculating the distance between pairs of sequences which is defined as the Euclidian distance between the feature matrices and (iv). clustering the sequences based on these distances using Agglomerative Hierarchical Clustering (AHC) algorithm.

We have followed the approach of the case study in [11] to implement their methodology on the Tennessee Eastman dataset. The average linkage criterion was used for the AHC algorithm. It was seen that it could classify the 42 sequences in the test dataset of the Tennessee Eastman process with 85.7% accuracy when full length sequences were given as input as compared to the 100% accuracy seen in the approach proposed in this paper. This could be due to the fact that their approach uses much less information for modelling and clustering for classification, unlike the HMM based methodology proposed here.

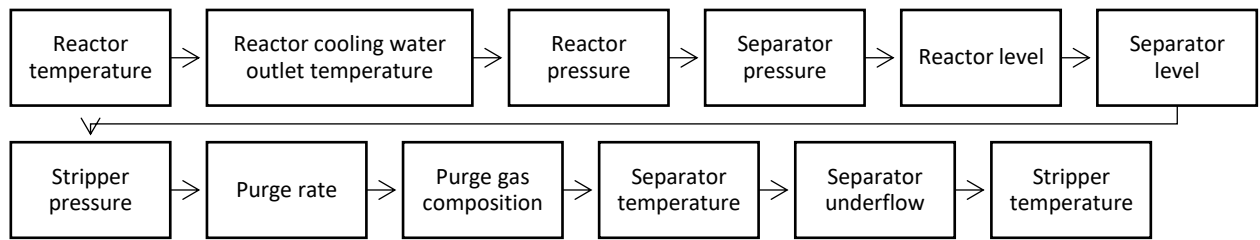

(a).

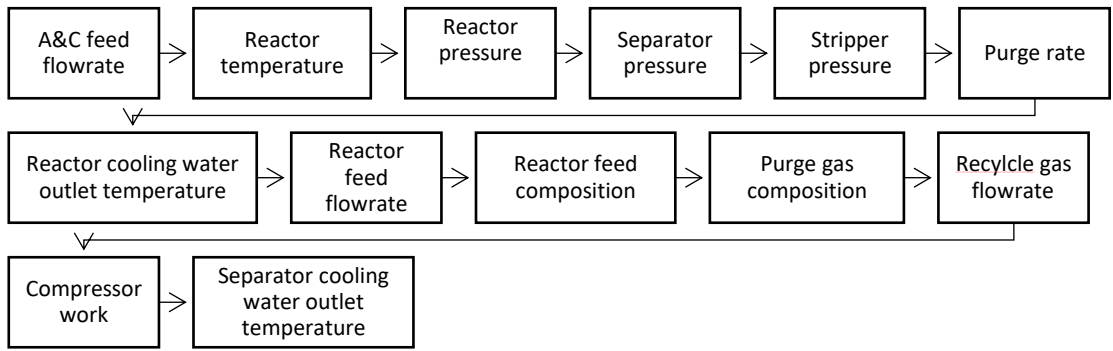

(b).

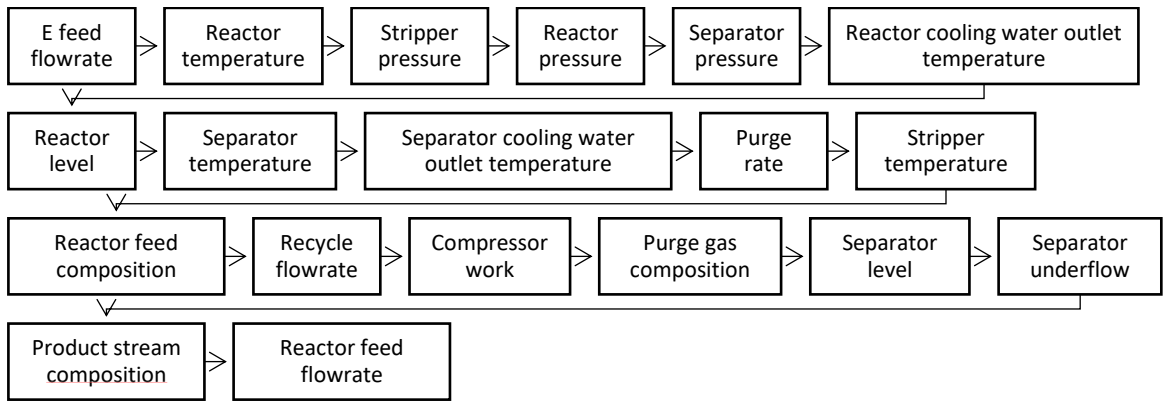

(c).

Figure 5: Fault propagation paths for (a). fault 1: reactor temperature gauge fault, (b). fault 2: C feed valve stiction and (c). fault 3: E feed valve stiction fault

## 4. CONCLUSIONS AND FUTURE WORK

An approach for diagnosing cause of alarm signals is proposed with the objective to guide the operator for early troubleshooting of the hazardous scenario. The paper proposes the application of a statistical model-based classification approach based on discrete hidden Markov models (HMM) to diagnose the fault using only binary alarm systems data as input. The key features of the diagnoser are its low computational complexity and reasonably interpretable model parameters. Other features of the diagnoser are that it allows integrating the information from process dynamic simulators, alarm system database as well as process safety review processes to correctly detect the probable cause of alarm signals. The diagnoser analyses the sequence in which alarms are triggered to identify the most and the second most probable cause. The doubly stochastic process allows to capture the effects of differences in propagation paths that arise due to stochastic elements like different fault magnitudes and random sensor noises. Information from failure rate databases can also be integrated through the HMM parameter $\pi$.

Application of the diagnoser on the Tennessee Eastman process showed that the model was successful in early classification of the fault with high accuracy. The proposed method was compared to the similarity approach of [11] that gave a lower accuracy. In the future, we will develop a diagnoser with multiple HMMs that can diagnose the cause of alarm signals with a guaranteed classification accuracy for a fixed length of input sequence. Another limitation of this approach is the assumption that faults occur one at a time. The possibility of having multiple faults at a time will also be investigated soon. Future work will also focus on building model-based alarm systems, using the concept of dynamic safety sets proposed in[64] to characterize process safeness.

Acknowledgments

Financial assistance from the Artie McFerrin Department of Chemical Engineering of Texas A&M University, Texas A&M Engineering Experiment Station and Mary Kay O'Connor Process Safety Centre is gratefully acknowledged.
Data availability statement

The data that support the findings of this study are available from the corresponding author upon reasonable request.